\documentclass[twocolumn,prd]{revtex4}
\usepackage[utf8]{inputenc}
\usepackage{graphicx}
\usepackage{setspace}
\usepackage{amsmath}
\usepackage{tensor}
\usepackage{color}

\newcommand\NRSurAuthors[1]{Blackman et al.}
\newcommand\FieldPRX[1]{Field et al.}

\begin{document}
\title{Cost Minimization in Acquisition for Gravitational Wave Surrogate Modeling}
\author{ Karl Daningburg }
\affiliation{ Rochester Institute of Technology }
\author{ Richard O'Shaughnessy }
\affiliation{ Rochester Institute of Technology }
\begin{abstract}
	Gravitational wave science is dependent upon expensive numerical simulations, which provide the foundational
        understanding of binary merger radiation needed to interpret observations of massive binary black holes. 
 The high cost of these simulations limits large-scale campaigns to explore the binary black hole parameter space.
 Surrogate models have been developed to efficiently interpolate between simulation results, but these models require a
 sufficiently comprehensive sample to train on.  Acquisition functions can be used to identify points in the domain for simulation. We develop a new acquisition function which accounts for the cost of simulating new points. We show that when applied to a 3D domain of binary mass ratio and dimensionless spins, the accumulated cost of simulation is reduced by a factor of about 10.
\end{abstract}
 
\maketitle

\section{Introduction}

The Advanced LIGO \cite{2015CQGra..32g4001T} and Virgo \cite{Acernese_2014} instruments are regularly observing the gravitational waves generated by the merger of two
compact objects \cite{DiscoveryPaper,LIGO-O2-Catalog,LIGO-O3-O3a-catalog,LIGO-O3-O3b-catalog}.
 To determine the parameters of the binaries responsible for these observations, each measurement is compared directly or
 indirectly with predictions for the gravitational radiation produced when binaries merge \cite{gw-astro-PE-lalinference-v1,gwastro-PENR-RIFT,gwastro-pe-bilby-2018}.   Numerical simulations provide the authoritative (and often only) foundational insight into radiation generated during the late stages of mergers \cite{2010RvMP...82.3069C}.   However,  these simulations are extremely expensive, making complete exploration of the domain
 infeasible. 

Some scientists have turned to surrogate modeling, a class of mathematical methods which emulate an underlying
function.  
Surrogate models have been widely applied to gravitational waveform modeling
\cite{gw-astro-ReducedOrderQuadraturePE-TiglioEtAl2014,2014PhRvX...4c1006F,gwastro-approx-ROMNR-Blackman2015,gwastro-pe-ROM-IMRPv2-2016,2019PhRvR...1c3015V,2017PhRvD..95j4023B,SurrogateGWs2014,gwastro-SEOBNRv4, gwastro-mergers-IMRPhenomP, gwastro-mergers-IMRPhenomXP, 2020PhRvD.102d4055O,2017PhRvD..96l3011D},
both to accelerate existing approximations and to interpolate between detailed numerical relativity simulations.  
Given a gravitational wave model evaluated on candidate binary parameters $x_k$,  a surrogate model interpolates
the waveform's asymptotic amplitude and phase across the binary parameters $x$ (and time).   However, to date these
calculations all adopt a simplified way to select the initial training points $x_k$: they are usually selected with a
greedy algorithm based on a \emph{fast approximation} to the true waveform \cite{2017PhRvD..95j4023B}, independent of the
cost of simulating those parameters $x_k$.

Our novel contribution is to incorporate both interpolator uncertainty and acquisition cost into our acquisition function. This approach balances the goals of (i) accurately modeling the objective function over the entire domain by picking points of maximum variance and (ii) minimizing total acquisition cost by minimizing the cost of selected points. Importantly, our goal is not to create a surrogate model which rivals existing surrogate models in accuracy or speed; it is only to demonstrate the potential of cost-weighted acquisition when it comes to training a surrogate model with Numerical Relativity as cheaply as possible.

This paper is organized as follows. In Section \ref{sec:method} we define the objective function we have chosen to
learn, then define acquisition functions to be tested.  We introduce a workflow for acquiring data and training a
Gaussian process. In Section \ref{sec:results} we examine the performance of the acquisition functions and resulting
surrogate models. In Section \ref{sec:conclude} we summarize the work and conclude with future prospects for this research.

\section{Method}
\label{sec:method}

\subsection{Surrogate review by example}

To be concrete, we introduce surrogate modeling for gravitational waves by summarizing the specific techniques adopted
by   \NRSurAuthors{} in  \cite{2017PhRvD..95j4023B} to generate a surrogate gravitational wave model for precessing binaries.
The complex dimensionless gravitational wave strain
\begin{gather}
	h(t,\theta,\phi;\boldsymbol\lambda) = h_+(t,\theta,\phi;\boldsymbol\lambda) - ih_{\times}(t,\theta,\phi;\boldsymbol\lambda)
\end{gather}
is expanded in terms of its polarizations $h_+$ and $h_{\times}$, where $t$ is time, $\theta$ and $\lambda$ are polar
and azimuthal angles of wave propogation from the binary system, and $\boldsymbol\lambda$ is the set of chosen
parameters to describe the system. \NRSurAuthors{} select black hole spin vectors and initial
mass ratio as parameters for their surrogate model; however, the method applies for any parameter of interest.   The angular dependence can
be efficiently represented with a spin-weighted spherical harmonic decomposition
\begin{gather}
	h(t,\theta,\phi;\boldsymbol\lambda) = \sum_{l=2}^{\infty} \sum_{m=-l}^l h^{lm}(t;\boldsymbol\lambda){}^{-2}Y_{lm}(\theta,\phi) %
\end{gather}
Following \NRSurAuthors{}, we will focus only on the $l=2,3$ modes, since higher-order modes have minimal impact on systematic error. 
Their surrogate models seek to represent each mode function $h_{lm}(t,\boldsymbol{\lambda})$ versus time and $\boldsymbol{\lambda}$.

\NRSurAuthors{}'s approach relies  on discovering a finite set of greedy parameters 
\begin{gather}
	G \equiv \{ \boldsymbol\Lambda _i \in \mathcal{T} \}_{i=1}^N 
\end{gather}
where $\mathcal{T}$ is a compact region of parameter space. They then perform NR simulations at each greedy parameter, resulting in the greedy solutions $\{W(t;\boldsymbol\Lambda _i)\}_{i=1}^N$. They transcribe these solutions into a regression model first by constructing an orthonormal linear basis $B_n = \{e^i(t)\}_{i=1}^N$ spanning the greedy solutions so that

\begin{gather}
	W(t;\boldsymbol \lambda) \approx \sum_{i=1}^n c_i(\boldsymbol \lambda)e^i(t)
\end{gather}
where coefficient $c_i$ is the inner product of $W(t;\boldsymbol\lambda)$ with $e^i(t)$. In principle the greedy parameters $G$ may be found by iteratively comparing the approximation to the true function $W(t;\boldsymbol\lambda)$ and selecting the point of maximum error:

\begin{gather}
	E_n(\boldsymbol\lambda) = \parallel W(\cdot ;\boldsymbol\lambda) - \sum_{i=1}^n c_i(\boldsymbol \lambda)e^i(\cdot ) \parallel
\end{gather}
However, due to the high cost of full numerical simulation, dense evaluation of this error is not feasible. \NRSurAuthors{} circumvent this by building a mock surrogate from a cheaper data source: post-Newton waveform models. They assume that the greedy parameters found in construction of the mock surrogate, $G^{PN}$, approximate the greedy parameters for the NR surrogate, and thus are able to select training data at a reasonable cost.

\begin{figure}
    \includegraphics[width=\columnwidth]{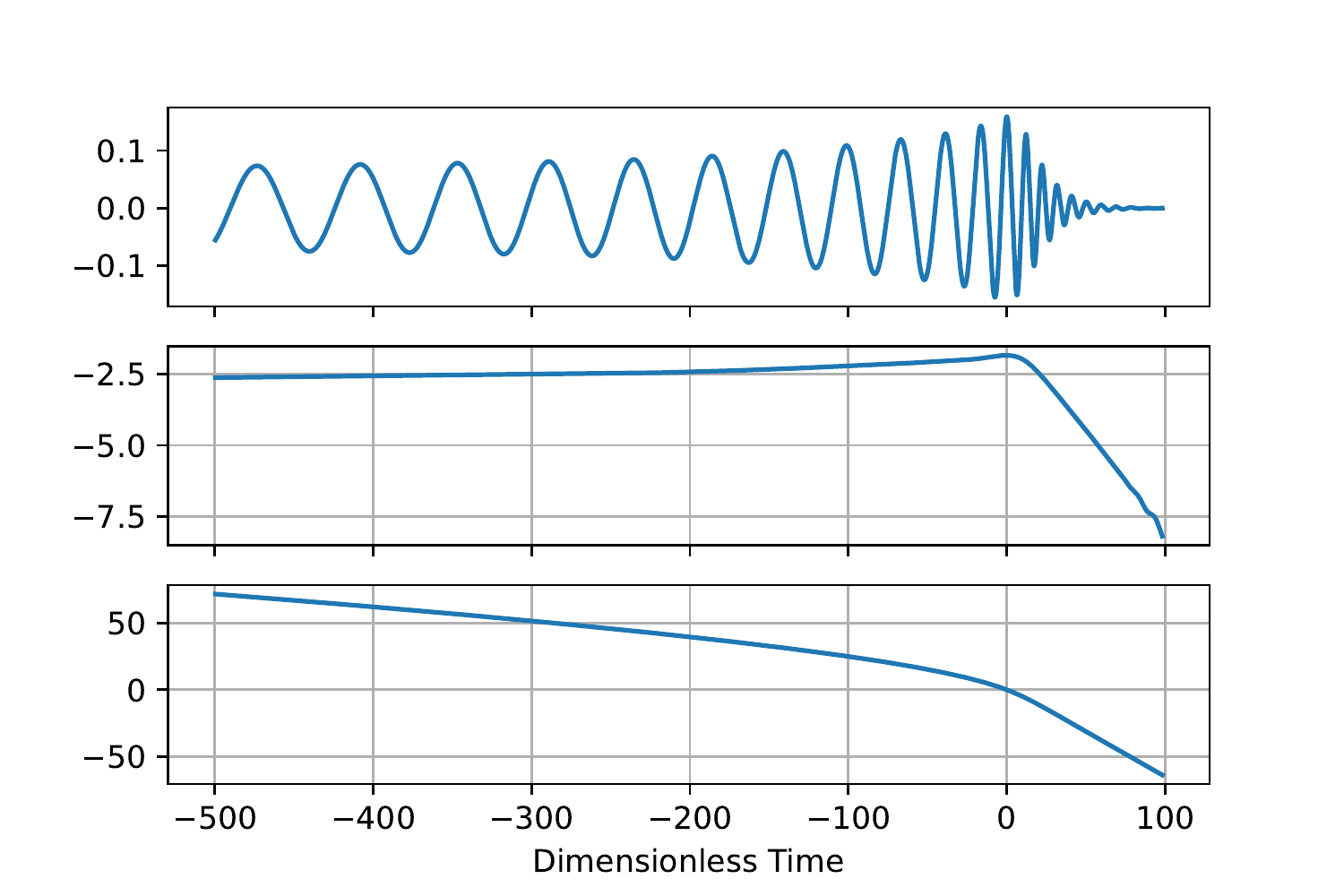}
    \caption{\label{fig:h_demo}Strain wave h(t) is depicted, as well as the wave's decomposition into log amplitude and phase. These smooth functions are more easily interpolated than the raw wave data and thus are used as input to the surrogate model.}
\end{figure}

Gravitational waves presented as functions of time have complicated dependence on time and binary parameters, which adds
additional artificial hurdles when attempting to interpolate them. A powerful technique is to decompose the waveforms obtained by NR simulations into data pieces which vary more slowly and thus are more easily modeled. These disparate models can be combined to form a single surrogate model. Commonly the phase and amplitude of the waves are used as the data pieces; however, for the precessing waveforms which \NRSurAuthors{} are targeting, a more sophisticated decomposition is needed. First, each waveform is translated into a coprecessing coordinate frame. Next, a Gaussian filter is used to remove the effect of nutation of the frame's rotational axis. Finally, the waveforms are decomposed into symmetric and antisymmetric amplitudes and phases
\begin{gather}
	A_\pm ^{l,m}(t) = \frac{1}{2}\left( \mid \tilde{h}^{l,m}(t) \mid \pm \mid \tilde{h}^{l,-m}(t) \mid \right) \\
	\varphi _\pm ^{l,m}(t) = \frac{1}{2} \left(  \varphi(\tilde{h}^{l,m}(t)) \pm \varphi (\tilde{h}^{l,-m}(t)) \right)
\end{gather}
where $\tilde{h}$ is the waveform in the coprecessing frame.  The surrogate training data and output are always
expressed using this coprecessing-frame  amplitude-phase decomposition.
Similarly, surrogate models can be constructed for the  precessional dynamics themselves, needed to translate between
the simulation and coprecessing frame.
Figure \ref{fig:h_demo} illustrates the rationale for using an amplitude-phase decomposition for the simple case of a
nonprecessing binary black hole's dominant mode: in this amplitude-phase decomposition, the two salient functions
involved in $h(t)$ are slowly varying in time (and versus binary parameters).

Now that amplitudes and phases for the decomposed waveforms have been extracted, interpolation on the data can be performed. \NRSurAuthors{} use an empirical interpolant via Singular Value Decomposition to unify the $\tilde{h}(t)$ waveforms into a single surrogate model $W_S(t;\boldsymbol\lambda)$. Despite relying on a SVD algorithm to find basis functions, \NRSurAuthors{} use a greedy algorithm to select time points over which to interpolate with these functions. For a detailed discussion on building reduced bases with SVD versus greedy algorithms, see Appendix B of \cite{2017PhRvD..95j4023B}. \NRSurAuthors{} use the so-called ``empirical time" algorithm developed in \cite{SurrogateGWs2014}. Given a reduced basis $\{e_i\}_{i=1}^m$ whose span approximates the objective function (in our case, gravitational wave phase and amplitude), the Empirical Interpolation Method (EIM) yields a set of $m$ empirical times $T_i$ chosen from the complete set of times $t_i$. These times are selected by comparing the value of the empirical interpolant 
\begin{gather}
	\mathcal{I}_j [h](t;\boldsymbol\lambda) = \sum_{i=1}^j C_i (\boldsymbol\lambda)e_i(t)
\end{gather}
to the basis functions evaluated at all times, where the $C_i$ coefficients are the solutions to the interpolation problem at the empirical times. At each step, one more basis function is added to $\mathcal{I}_j [h](t;\lambda) $; the time of maximum error is found and added to the set of empirical times $T_i$. This process continues until the number of times chosen is equal to the number of basis functions $m$.

\NRSurAuthors{} acquire data for their surrogate model by maximizing error between the true function and the surrogate. Since NR data is far too expensive to obtain in the quantities needed for this method, they evaluate a cheaper objective function to acquire new points. Another type of acquisition function which may be more easily used with an expensive reference function is one based on minimizing uncertainty rather than error. This formulation of the interpolation problem lends itself to Gaussian processes, which provide quantification of uncertainty throughout the domain. Therefore, we propose to focus on Gaussian processes as interpolators and measures of uncertainty as a main ingredient in our acquisition functions.

\subsection{Gaussian process interpolation}

Stochastic processes may be understood as an extension of probability distribution functions from random variables to functions. We utilize the Gaussian process (GP), which by generalizing the Gaussian distribution simplifies computations required for learning \cite{book-Rasmussen-GP}. Analogous to the mean and variance parameters of the Gaussian, GPs are fully defined by a mean function, $\mu(x)$, and a covariance (or kernel) function, $k(x, x')$. GPs can interpolate across a domain, representing the infinite number of possible true functions as variance. Once trained, a GP can be evaluated to provide new data in emulation of the objective function. 
Gaussian processes have been previously employed to perform the interpolation needed to construct surrogate models and
identify candidate points for followup analysis; see, e.g., \cite{2017PhRvD..96l3011D}, which is similar to our
cost-neutral approach.

The choice of a kernel function is not a straightforward one. Since kernel functions encapsulate assumptions about the function to be learned, prior knowledge of this function is useful. For this introductory work, we use the simple and commonly-used squared exponential (SE) function, with the understanding that further investigation is worthwhile:

\begin{gather}
	K(X,Y) = \sigma^2 \exp \left\{ - \frac{(X-Y)^2}{2l^2} \right\}
\end{gather}
where $X$ and $Y$ are the two inputs, $\sigma^2$ is the variance, and $l$ is the characteristic length scale, which modifies the correlation between nearby data points. The SE function has the advantage of being easy to evaluate and infinitely differentiable, and it yields smooth interpolation between data points. 

Gaussian processes are inherently probabilistic, allowing them to express uncertainty in a given domain as \textit{variance}. The equation for a GP's variance is shown below, where $K()$ represents the selected kernel function, $X$ represents the training data, and $X_*$ is the point at which variance is computed.
\begin{gather}
	V = K(X_*, X_*) - K(X_*,X)[K(X,X)]^{-1} K(X,X_*) \label{eqn:GP_var}
\end{gather}

\subsection{Simulation Cost}

Although Numerical Relativity (NR) simulations were not used for training this surrogate model, we develop this model with the intent of directly targeting and interpolating NR simulations.  As our goal is a reduction of
simulation and training  cost, we develop a cost function to estimate the expense of simulating points in the domain.
Our colleagues estimate \cite{NR_cost} the length of a simulation scales approximately with mass ratio $q$, as does the
resolution required to maintain accuracy; therefore the cost dependence on mass ratio can be modeled as
$q^2$. Meanwhile, dimensionless spins of the two bodies have very little effect on the simulation cost until spin of 0.6
is reached. Thereafter, the cost increases rapidly and diverges as spin approaches 1. 
After consulting with an NR group \cite{NR_cost}, the following piecewise function models the cost-spin relationship:

\begin{subequations}
\label{eq:cost}
\begin{gather}
    {\cal S}(\chi)= \begin{cases} 
      1 & \chi\leq 0.6 \\
      \frac{0.4}{1-\chi} & 0.6 < \chi \leq 0.9 \\
      \frac{0.04}{(1-\chi)^2} & 0.9 < \chi < 1
      \end{cases}
\end{gather}
For our chosen domain of mass ratio and spins for bodies 1 and 2, the cost function is as follows:

\begin{gather}
	C(q, \chi_1, \chi_2) = q^2 + {\cal S}(\chi_1) + {\cal S}(\chi_2) \label{eqn:cost}
\end{gather}
where $q$ is mass ratio $\frac{m_1}{m_2}$, $m_1 > m_2$, and $\chi_1$ ($\chi_2$) is dimensionless spin magnitude for body 1 (2). %
\end{subequations}

\subsection{Acquisition}
\label{sec:acquisition}
Acquisition refers to the collection of a data point, whether by experimentation or simulation. In order
to train our surrogate model most efficiently, we develop acquisition functions which assign a value to each point in
the domain. Optimizers are used to maximize these acquisition functions, and the resulting point in the domain is
selected for the next simulation.

Abstractly, we wish to formulate an acquisition function which balances the need to find points of maximum uncertainty with the requirement of keeping simulation costs at a minimum. Having defined variance and cost in Equations (\ref{eqn:GP_var}) and (\ref{eqn:cost}), we can now construct an acquisition function which quantifies our learning per unit cost:
\begin{gather}
	A = \frac{V}{C}
\end{gather}
Maximizing this function at each step constitutes a greedy algorithm for learning as much about the objective function as possible for the lowest cost.

For the purpose of comparison, we test both a variance maximizing routine (see, e.g., \cite{2017PhRvD..96l3011D}) and a variance-to-cost ratio maximizing routine. For brevity, we will refer to these as $V$ and $V:C$ routines, respectively.

\subsection{Time Selection}
\label{sec:TimeSelection}
The data available for a given point in parameter space is comprised of a complete gravitational waveform which may be expressed in the frequency domain or the time domain. It is desireable to select an optimal set of times (or frequencies) at which to measure the amplitude or phase of each available GW and interpolate that data through parameter space, allowing us to reconstruct any waveform in the domain.

The algorithm for time selection proceeds as follows.
Given a set of GWs associated with a set of parameters, we seek a minimal subset of dimensionless time points $\tau$ to fully characterize the behavior of any waveform in the domain; we will therefore call this set \textit{characteristic time} $\tau_c$.
To initialize the algorithm, five evenly-spaced time points are selected. The data
corresponding to these times are fed to SciPy's interp1d routine, which trains a cubic spline interpolator for each available GW. We use this spline to interpolate across 50000 evenly spaced time points from -500 to +75 $\tau$. The root-mean-square error
\begin{gather}
	\text{RMSE}\left( \phi_j \right) = \frac{1}{n}\sum_{i=1}^n \sqrt{\left(\frac{\phi_{f}^{(ij)} - \phi_{c}^{(ij)}}{\max |\phi_{f}^{(i)}| }\right)^2}
\end{gather}
is computed at each time step across all splines. Here $\phi$ is the GW phase, $j$ indicates one of the 50000 time steps, $n$ is the number of GWs available, $\phi_{f}$ is the fiducial spline, and $\phi_{c}$ is the spline trained only on $\tau_c$. The time of maximum error is chosen to be added to the spline training data. The process iterates until the maximum $\text{RMSE}\left( \phi_j \right)$ falls below a desired threshold. These characteristic time points and the corresponding values of phase or amplitude are used to train Gaussian processes for interpolation across parameter space. New GWs can be constructed by training new splines on the data interpolated by the GPs. 

\subsection{Error Metric}

We can evaluate the model's performance at any point in the space $\mathcal{T}$ by comparing it to the source of our
data, IMRPhenomD. A brief discussion of \textit{mismatch}, a common metric for quantifying the difference between
waveforms, is necessary. Given two waveforms, a fiducial wave $h_1$ and a surrogate approximation $h_2$, we require an
objective measure of how well $h_2$ approximates $h_1$. For a single mode we compute the  \textit{match} \cite{gwastro-mergers-HeeSuk-FisherMatrixWithAmplitudeCorrections}:
\begin{gather}
	\mathcal{O} =  \frac{\max_t \left |\int_{-\infty}^{\infty} h_2^*(f) h_1(f) e^{i2\pi f t} df \right |}{||h_1||\;  ||h_2||}
	\label{eqn:overlap}
\end{gather}
This expression  is evaluated using 
a white-noise detector power spectrum; alternatively, this expression corresponds to the  standard equally-weighted
Hilbert space inner product in time or frequency.
 Empirically we estimate the overlap
from our discrete data using an inverse Fourier transform
\begin{gather}
	\mathcal{O} =\max_t \text{IFT}\left[ N \left( \frac{\tilde{h}_2^*}{\left\lVert\tilde{h}_2\right\rVert } \right) \left( \frac{\tilde{h}_1}{\left\lVert\tilde{h}_1\right\rVert } \right) \right]
	\label{eqn:overlap:approx}
\end{gather}
where N is the number of samples, $\tilde{h}$ indicates a single wave mode in the frequency domain, and * represents the complex conjugate. Two waves identical to a constant factor will have an overlap of unity; therefore to reframe as an error metric we compute the mismatch as
\begin{gather}
	\mathcal{E} = 1 - \mathcal{O}
	\label{eqn:mismatch}
\end{gather}
 We chose this approach since this surrogate model is intended to be generic, independent of any detector's power spectrum. For a more complete discussion of waveform mismatch, see \cite{2017PhRvD..95j4023B}. 

\subsection{Testing with fast aligned model}

Our surrogate modeling methods are similar in principle to those outlined in \cite{2017PhRvD..95j4023B}. Given a physical system parametrized by $\boldsymbol\lambda \in \mathcal{T} $, where $\mathcal{T}$ is a compact region in parameter space, we seek cheaply evaluated functions of time $W(t;\boldsymbol\lambda)$ that describe the system. For our model $\boldsymbol\lambda$ includes mass ratio $q$ and spins ($\chi_{1z}$ and $\chi_{2z}$) of black holes in a binary black hole system and $\mathcal{T}$ encompasses a range of mass ratios and spins for which an existing model is available.

Although NR is the eventual data source for this type of model, we begin with a more rapidly solved method: IMRPhenomD \cite{IMRPhenomD}. IMRPhenomD is a phenomenological model which evaluates the gravitational wave signal of black hole binaries throughout the inspiral, merger, and ringdown phases. This model is capable of producing data for non-precessing black holes of mass ratio up to 18:1 and dimensionless spin of 0.85, or 0.98 for 1:1 mass sytems. IMRPhenomD is a hybrid model which incorporates both Effective-One-Body data and NR simulations for tuning. It has demonstrated error of under 1\% versus NR test waveforms \cite{IMRPhenomD}. We selected this model to enable more rapid training of our surrogate model and thorough error analysis of the methods developed. Its close approximation of NR results ensure translation of our methods to NR data acquisition with similar performance.

Once our parameter space is populated following the method outlined in Sec. \ref{sec:acquisition}, we turn to the task of interpolating the data across both time and the parameter space. We seek a set of dimensionless times $\tau_0,\ldots,\tau_k$ for each value to be interpolated across parameter space which characterizes the behavior of these values over time for all systems in the target region $\mathcal{T}$. Using the method described in Sec. \ref{sec:TimeSelection}, we choose characteristic times $\tau_c$ for both phase and amplitude of the GWs. Importantly, the times for phase and the times for amplitude are selected separately, since the behavior of these two values is disparate. Once the characteristic times are selected, Gaussian processes are trained on the value of amplitude or phase at each characteristic time for all available gravitational waves. 

To solve the surrogate model at any point in parameter space, first we evaluate the Gaussian processes at all characteristic times for both amplitude and phase. Then we reconstruct the full amplitude and phase curves in time using the natural cubic spline interpolant. These curves can be used to reconstruct the original strain waveform $h(t)$. To get a succinct understanding of competing models' performance, we evaluate the wave mismatch [see Eqs. (\ref{eqn:overlap}) and (\ref{eqn:mismatch})] at select points distributed evenly through parameter space and take the L2 norm to produce a single performance metric. We compute this error metric at strategically chosen intervals of acquisition to understand the overall error trend as more points are acquired.

\section{Results}
\label{sec:results}

\begin{figure}
    \includegraphics[width=\columnwidth]{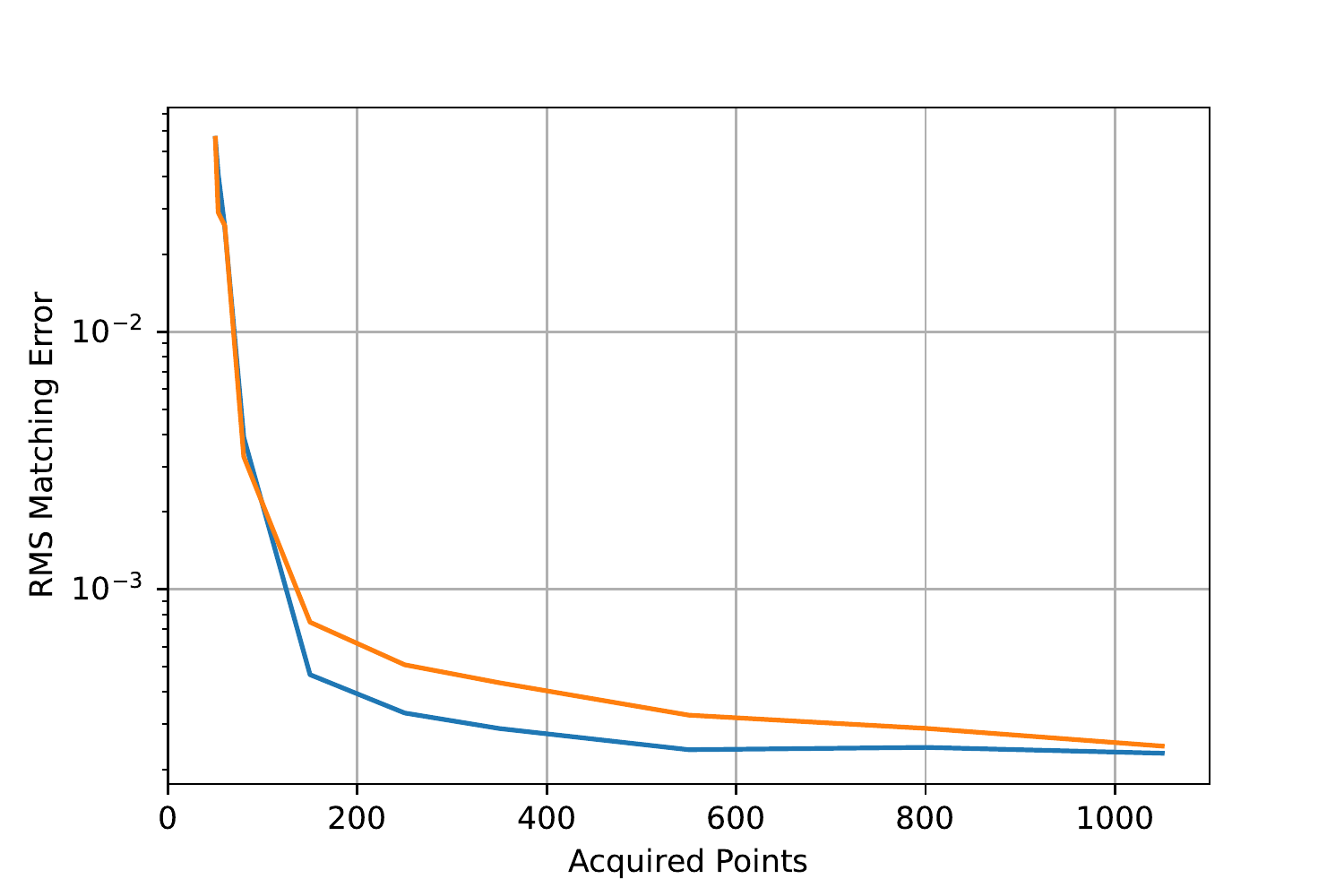}
    \caption{The mismatch (\ref{eqn:mismatch}) of two surrogate models over data acquisition. In blue the model acquires new points based only on GP variance; in orange, the ratio of variance to cost. The variance-to-cost acquisition function remains competitive with the variance alternative in all but the most high-cost areas.}
	\label{fig:error_vs_pts}
\end{figure}

To compare the performance of two acquisition functions of (a) variance maximization (hereafter $V$) and (b) variance-to-cost ratio maximization (hereafter $V:C$), we devised the following test. For parameter space $\mathcal{T}$ we chose the three dimensions upon which simulation cost relies: mass ratio $q$ and the spins $\chi_1$ and $\chi_2$. Mass ratio $q$ we allowed to vary from 1 to 5, while each $\chi$ ranged from 0 to 0.99. For the purposes of characteristic time, we chose a maximum L2 norm error of 0.0001 across all acquired waveforms, with a maximum number of characteristic time points of 200. Recall that for each characteristic time point, a Gaussian process will be trained on the data from that moment in time for all acquired waveforms (see subsection \ref{sec:TimeSelection}); therefore the cap of 200 is chosen to keep runtime reasonably low. We initialize the model with 50 data points selected randomly through Latin Hypercube Sampling, and then begin the iterative process of training and acquiring new data.

\subsection{Global performance}
A comparison of the two surrogate models is presented in Figure \ref{fig:error_vs_pts}. For this figure, the mismatch was measured at 125 points evenly spaced in the three dimensional parameter space. Mass ratio $q$ was sampled in intervals of 1 from 1 to 5, and spins $\chi_1$ and $\chi_2$ were sampled in intervals of 0.2 from 0 to 0.8. These samples match the limits of the parameter space with the exception of $\chi$, which extend to 0.99. We removed $\chi = 0.99$ from the global performance calculation due to the relatively high error in this expensive region dominating the L2 error norm of the $V:C$ routine. By the metric of mismatch vs number of acquired points, the $V$ scheme slightly outperforms the $V:C$ scheme until the two converge after 1000 acquisitions.

\begin{figure}
    \includegraphics[width=\columnwidth]{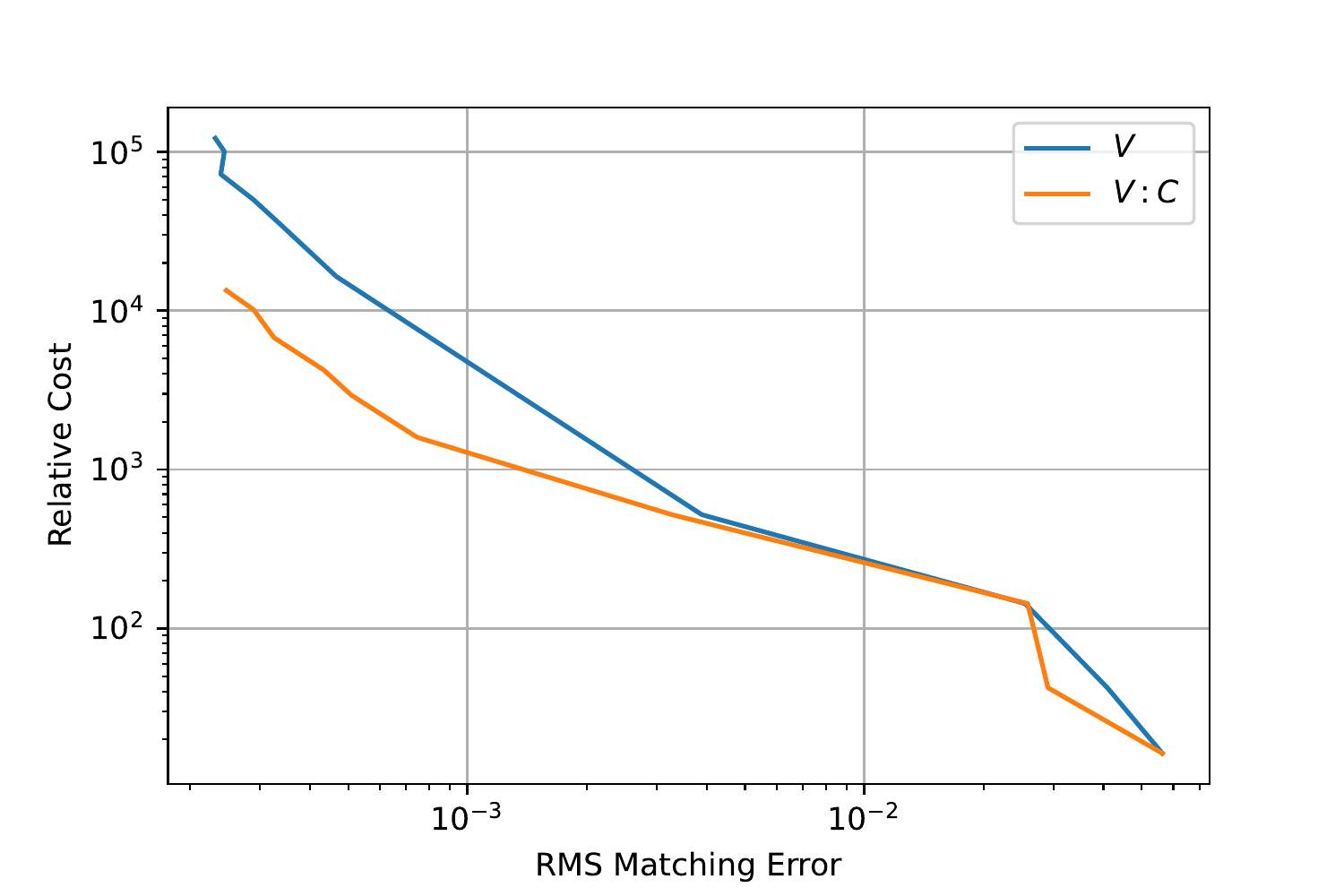}
    \caption{The relative acquisition cost of two surrogate models over RMS mismatch [Eq, (\ref{eqn:mismatch})]. In blue the model acquires new points based only on GP variance; in orange, the ratio of variance to cost. The $V:C$ acquisition function achieves the same RMS match error across the majority of the parameter space with nearly an order of magnitude smaller cost once error is driven below $10^{-3}$.}
	\label{fig:error_vs_cost}
\end{figure}

Reframing the problem as maximizing learning per unit cost rather than per data point acquired reveals the advantage of the $V:C$ acquisition function. Figure \ref{fig:error_vs_cost} shows the relationship between matching error and relative cost of acquired data points. While performance per cost is similar at higher errors, as the error is driven lower by further data acquisitions the $V:C$ routine strongly outperforms the $V$ routine. The $V:C$ routine achieves the same RMS matching error across the described domain for about an order of magnitude smaller cost.

\subsection{Performance over parameter space}
\begin{figure}
    \includegraphics[width=\columnwidth]{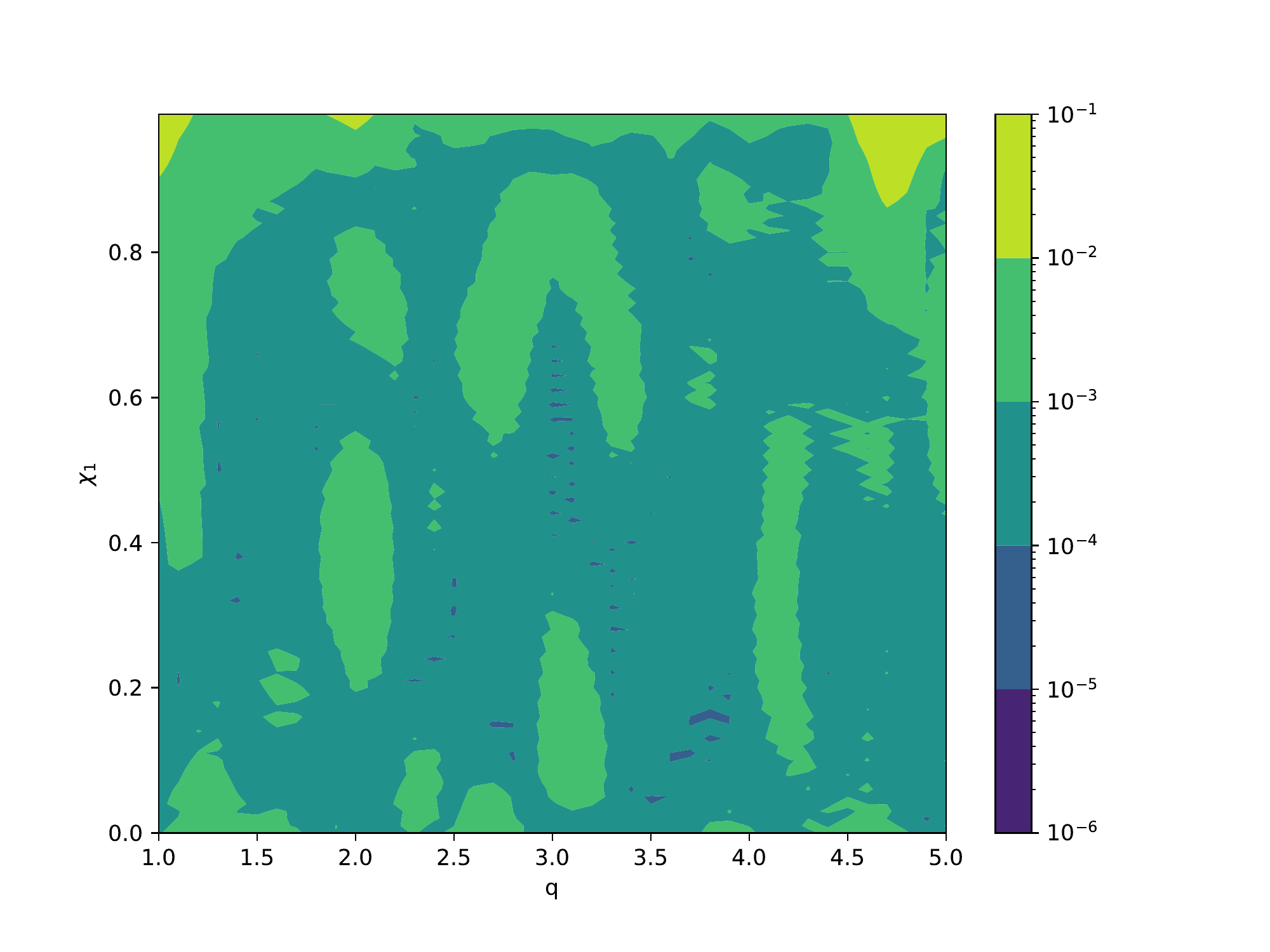}
    \caption{The mismatch in a slice of parameter space for the $V:C$ routine after 1000 acquisitions. Note that error remains high in a few ultra-high cost regions. For this slice of parameter space, $\chi_2$ is set to 0.99.}
	\label{fig:paramspace_error}
\end{figure}

The $V:C$ methods achieve their improved global performance by sacrificing exploration and thus accuracy in the most
costly regions.  To highlight this tendency, Figure \ref{fig:paramspace_error} shows the mismatch versus binary
parameters after a fixed number of acquisitions.  While the $V:C$-derived surrogate model works extremely well in
low-cost regions, it does not as reliably reproduce high-cost waveforms. 

To date, nature has provided principally low-spin compact binaries, so a focus on low-spin exploration seems warranted. 
However,  high-spin binaries have considerable discovery potential and interesting physics.  A more thorough review of
the science return of a given GW model based on its accuracy in different regions of parameter space is required to
better quantify this tradeoff.

\subsection{Discussion}

There are several key distinctions between the surrogate model developed here and the model developed by \NRSurAuthors{}
\cite{2017PhRvD..95j4023B}. 
Perhaps the most important distinction is the use of a Gaussian process as an interpolant rather than an SVD-based
empirical interpolant. Using a Gaussian process opened up the possibility of using the interpolator's uncertainty
(variance) about the objective function in order to place simulations to acquire more data efficiently.   
This approach is simpler and more direct than the method used in \cite{2017PhRvD..95j4023B}; \NRSurAuthors{} had to
resort to building a mock surrogate to evaluate error between the interpolator and the data source, since NR simulations
are too expensive for this type of acquisition. This complicates the model training process, and the use of variance as
a stand-in for error circumvents this step and simplifies our algorithm.
However, as a non-spectral method, the specific  GP interpolation adopted in this work converges relatively slowly
versus the number of new simulations acquired.  
We expect that our technique can be applied to scenarios with other empirically-estimated error metrics like the SVD
approach.

The second most important difference between the surrogate described in \cite{2017PhRvD..95j4023B} and the model
presented here is incorporation of simulation cost. While both approaches seek to minimize the total
number of NR simulations required to train a model, our method gives consideration to the
variability in cost of the simulations themselves. While we did not directly compare data acquisition cost between
\NRSurAuthors{}'s model and our $V:C$ model, we did use the variance maximization routine $V$ as a stand-in for this
comparison. We found that a cost savings of about tenfold was gained by using the $V:C$ acquisition function.  
Of course, due to its extremely rapid (exponential) convergence, the asymptotic behavior of a spectral interpolation
should be superior.  Our goal in this study is simply to demonstrate the utility of incorporating cost estimates into
surrogate simulation targeting.  

\section{Conclusion}
\label{sec:conclude}

In this paper, we demonstrate how to incorporate simulation cost into a concrete strategy for iteratively building a
surrogate for gravitational waves generated by black hole binaries.  By minimizing the estimated (Gaussian process) variance per
unit cost, we show how to target synthetic GW simulations to iteratively assemble a comparable-accuracy surrogate at
lower cost than a comparable cost-neutral approach.  At the mismatch scales appropriate to contemporary gravitational
waveform modeling (i.e., $10^{-3}$), we demonstrated our method  should be about an order of magnitude less costly than
a comparable cost-neutral approach over the same parameter space.  

Though we focused on synthetic gravitational wave simulations specifically, our methods could be directly transferable
to other data-sparse domains with high-cost simulations, including as a wide variety of applications in astrophysics
involving compact object mergers.

\bibliography{references,gw-astronomy-mergers,mybib,LIGO-publications,gw-astronomy-mergers-approximations,textbooks}

\begin{acknowledgements}
ROS is supported by NSF AST-1909534 and PHY 1912632, 2012057.  
We thank Hong Qi and Zoheyr Doctor for helpful feedback.
\end{acknowledgements}

\appendix
\section{Equations}

The Jacobian of the variance of a Gaussian Process using a squared exponential kernel was derived for use in a Python optimization routine:
\begin{gather}
	\begin{split}
		J_m(X_*) = \frac{1}{l^2}\left\lbrace \left( K^{-1} + (K^{-1})^T \right) K_* \right\rbrace^T \\ K_* \left(X_*^{(m,:)} - X^{(m,:)} \right)
	\end{split}
\end{gather}
where $J_m$ is the m\textit{th} element of the Jacobian, $K$ is the kernel evaluated on the training data, $K(X,X)$, $K_*$ is the kernel evaluated at a given point, $K(X_*,X)$, and the superscript $(m,:)$ represents the entire m\textit{th} row of a given matrix.

\end{document}